\documentclass[fleqn,10pt]{article}

\usepackage{fancyheadings}
\usepackage{amssymb}
\usepackage{latexsym}
\usepackage{epsfig}
\usepackage{alltt}

\usepackage{graphicx}
\usepackage{setspace}

\newcommand{\CL}{\textsf{Common~Lisp}}
\newcommand{\CLAZY}{\textsf{CLAZY}}

\newcommand{\code}[1]{\texttt{#1}}

\newcommand{\PLL}{$\mathbf{\mathsf{<\mathrm{LLP}\;\;\Pi\Lambda\Lambda>}}$}

\title{
\LARGE{\bfseries \CLAZY{}: Lazy Calling for \CL{}}}

\author{
Marco Antoniotti\\
\small Lab Linguaggi Programmazione -- Programming Language Lab \PLL{}\\
\small Dipartimento di Informatica, Sistemistica e Comunicazione,\\
\small Universit\`{a} Milano Bicocca\\
\small U14 - Viale Sarca 336, I-20126 Milan, \textsc{Italy}}

\date{}

\begin{document}

\pagestyle{empty}

\maketitle

\begin{abstract}
This document contains a description of a \CL{} extension that allows
a programmer to write functional programs that use \emph{normal order}
evaluation, as in \emph{non-strict} languages like Haskell.  The
extension is relatively straightforward, and it appears to be the
first one such that is integrated in the overall \CL{}
framework\footnote{A version of this note was presented at the
  1st European Lisp Symposium 2008, Bordeaux, France}.
\end{abstract}

\rhead[]{\textit{\CLAZY{}: Lazy Calling\ldots}}
\lhead[]{\textit{Antoniotti, M.}}
\chead[]{\PLL}
% \rfoot[]{}
% %\lfoot[\small\sf PHS 398 (Rev. 09/04)]{\small\sf PHS 398 (Rev. 09/04)}
% %\lfoot[DRAFT]{DO NOT DISTRIBUTE}
% \cfoot[]{}
% \cfoot[\small\sf Page \thepage]{\small\sf Page \thepage}

\pagestyle{fancy}

\section{Introduction}

\CL{} is a functional language (and also an imperative,
object-oriented one, which, moreover, can be used in a declarative
fashion).  As a functional language it falls in the category of
\emph{strict} languages like ML \cite{milner97:_def_standard_ml} and
OCaml \cite{leroy14:_ocaml}, unlike Haskell \cite{Haskell2010}, which
is in the category of \emph{normal-order} or \emph{lazy} languages.

That is to say that the following code will enter an infinite loop,
should it be executed at the \CL{} prompt.
\begin{alltt}

    cl-prompt> \textbf{(defun si (condicio ergo alternatio)
                  (if condicio ergo alternatio))}
    \textit{SI}

    cl-prompt> \textbf{(si t 42 (loop))}
\end{alltt}
In a \emph{lazy} language the function \code{si} (\code{if}
in Latin) would return 42 instead of waiting for the form
\code{(loop)} to produce a value.

In a bout of Haskell envy, I decided to look into some extensions to
\CL{} that would introduce ways to program in a lazy way. The result
may sound \emph{crazy}, and, in fact, a little bit it is.

The notion of \emph{lazy evaluation} dates back to the Algol days and
the notion of \emph{by-name} parameter passing.  In the Lisp camp, the
best known way to introduce a form of lazy evaluation is to implement
\emph{streams} as described in \emph{Structure and Interpretation of
  Computer Programs} (SICP) \cite{SICP}; incidentally this form of
lazy evaluation is also used by Okasaki \cite{okasaki98:_pfds} in his
exposition of \emph{functional data structures} in ML.

In SICP, streams are implemented using two primitives, \code{force}
and \code{delay}, which can then be used to build a lazy container
(the ``stream'') using a \emph{macro} \code{cons-stream}, and two
accessors \code{head} and \code{tail}.  A sufficient implementation in
\CL{} is the following:
\begin{alltt}
    (defmacro delay (expr) `(lambda () ,expr))

    (defun force (thunk) (funcall thunk))

    (defmacro cons-stream (head tail) `(cons ,head (delay ,tail)))

    (defun head (lazy-stream) (car lazy-stream))

    (defun tail (lazy-stream) (force (cdr lazy-stream)))
\end{alltt}
At this point there are several \CL{} packages floating around the
net, that implement this flavor of lazy evaluation. E.g., \code{Heresy}
\cite{lamari07:_heresy}, \code{funds} \cite{baine07:_funds} and \code{FSet}
\cite{burson07:_fset} are exemplars of this approach. \CLAZY{} goes
off a (different) tangent and provides a more fundamental way to build
such lazy constructions.

\subsection{Limits of the \code{delay}/\code{force} Duo}

Given \code{delay} and \code{force}, one could always implement the
operator \code{si} as a macro using \code{delay}, as in
\begin{alltt}

    (defmacro si (condicio ergo alternatio)
       `(if (force ,condition)
            (force (delay ,ergo))
            (force (delay ,alternatio))))

\end{alltt}
but this is a bit unsatisfactory as far as Haskell envy is
concerned. \code{si} cannot be funcalled in any meaningful way and
cannot be passed around as we would expect a regular function to
be.  A different solution is needed.

\section{Defining and Calling Lazy Functions}

It is possible to come up with a more satisfactory solution that will
allow us to bypass \code{delay} and \code{force}, at the price of
tweaking the ``calling convention''.  Then we can write \code{si}
as:
\begin{alltt}

    (\textbf{deflazy} (condicio ergo alternatio)
       (if condicio ergo alternatio))

\end{alltt}
where \textbf{\code{deflazy}} defines both \emph{lazy} and
\emph{strict} versions of the operator.

%\newpage
\noindent
The \emph{lazy function} \code{si} can now be called as
\begin{alltt}

    CL prompt> \textbf{(\textit{lazy:call} #'si t 42 (loop))}
    \textit{42}

\end{alltt}
I.e., \textbf{\code{lazy:call}} is the lazy version of \code{funcall}.
The complexity of writing lazy code is thus moved to the call points.
This may or may not be desirable, but it can be argued that this is a
slightly better way than having to manually \code{force} expressions.
In any case, the \CLAZY{} approach still uses the
\code{delay}/\code{force} duo under the hood, and they are available
for more manual intervention.

From the example above, it should be apparent that \code{lazy:call} is
a macro that does something special with the call, recognizing
functions that are defined via \code{deflazy}.  As a matter of facts,
the expansion of \code{lazy:call} looks like this:
\begin{alltt}

   (lazy:call <op> <arg1> <arg2> ... <argN>)
   \(\Longrightarrow\)
   (funcall <\textit{lazyfied} op>
            <\textit{thunked} arg1>
            <\textit{thunked} arg2>
            \ldots
            <\textit{thunked} argN>)

\end{alltt}
The ``lazy'' version of \code{<op>} is defined by
\code{deflazy} and each \code{<\textit{thunked}~arg$_i$>} is a closed
over version of the argument as if \code{delay} was invoked on it.

Of course, a simple version of such idea can be easily implemented
with a few macros, however, a well integrated version within the
overall \CL{} environment requires a few more bits and pieces.  As
example, \CLAZY{} wants to make the analogy between \code{lazy:call}
and \code{funcall} as tight as possible.  This means that we need a
way to pass (almost) regular \code{lambda}'s to \code{lazy:call}.
This can be done the special operator \code{lazy}, which acts as
\code{function}; moreover, it does wrap around the \code{function}
operator as expected.  See Figure~(\ref{fig:lazy-special-op-use}) for an
example.

\begin{figure}
\hrulefill
\begin{alltt}
   CL prompt> (lazy:call (\emph{lazy} #'(lambda (condicio ergo alternatio)
                                    (if condicio
                                        ergo
                                        alternatio)))
                         t
                         (+ 20 20 2)
                         (loop))
   \textit{42}
\end{alltt}
\hrulefill
\caption{An example of the use of the special operator \code{lazy}.}
\label{fig:lazy-special-op-use}
\end{figure}

Extra work is needed to handle \code{\&optional} and \code{\&key}
parameters, but the overall design lies in this tweaking of the
calling point and in allowing lazy functional objects to be passed
around as regular functions (of course to be called via
\code{lazy:call}).

\subsection{Example: Lazy Functional Conses}

Another example which turns out to be more easily realizable with
\CLAZY{} is the standard ``\emph{conses are functions}'' one.
\begin{alltt}

   (deflazy conc (head tail)
      (lambda (selector)
          (ecase selector
            (car head)
            (cdr tail))))

   (deflazy head (cons) (funcall cons 'car))

   (deflazy tail (cons) (funcall cons 'cdr))

\end{alltt}
Now, we can build truly lazy lists\footnote{Note where the
  \code{(loop)} calls appear.}
\begin{alltt}

   CL prompt> (defparameter ll
                 (lazy:call 'conc
                            1
                            (lazy:call 'conc
                                       (loop)
                                       (lazy:call 'conc
                                                  3
                                                  (loop)))))
   \textit{LL}

   CL prompt> (head (tail (tail ll)))
   \textit{3}

\end{alltt}
Or the usual streams from SICP as the \code{integers} here below.
\begin{alltt}

   (defun integers-from (n)
      (lazy:call 'conc n (integers-from (1+ n))))

   (defparameter integers (integers-from 0))

\end{alltt}
Yet, it must be noted that having normal order evaluation at one's
disposal naturally leads to the implementation of much more complex
and sophisticated functional software, as in the case of the
integrators in Section~3.5 of \cite{SICP}.

\subsection{Extra Considerations}

\CLAZY{} is supposed to be used in a very controlled way.  While it is
true that it adds \emph{normal order evaluation} to \CL{}, the user
must remember that s/he is not using Haskell or a similar language.
At its core, \CL{} is a \emph{strict} language, which allows
side-effects; not a good mix to produce lazy code in a careless way.
See also the note on \emph{normal order evaluation} in Section~3.5 on
streams of \cite{SICP}.

\section{Reference Implementation}

The \CLAZY{} reference implementation can be found
at \texttt{common-lisp.net}.
The implementation lies within a package nicknamed \code{LAZY} and
is based on the macros \code{lazy:call}, \code{lazy:deflazy}, and
\code{lazy:lazy}.

The \code{lazy:call} macro is used at calling time (as the name
implies).  The \code{deflazy} macro is used to define functions.
The \code{lazy} ``special operator'' returns a functional
object that should be called in a lazy way, although the system is set
up in such a way to ``pass through'' constant values (as
tested by \code{constanp}).

The reference implementation is based on the pre-processing of lambda
list arguments by \code{deflazy}: each argument is substituted by an
internal name, which is expected to be bound to a \emph{thunk}
generated by \code{lazy:call} as per \code{delay}.  In the body of a
lazy function (or of a \emph{lazy lambda}) each lambda list argument
is actually re-defined as a a \code{symbol-macrolet}, which expands in
the appropriate \code{force} call.  \code{deflazy} installs the lazy
version of the function being defined in the property list of the
function name.

\paragraph{Ordinary Lambda List Processing.} As noted before, \CLAZY{}
pre-processes \code{\&optional} and \code{\&key} arguments in such a
way to preserve the expected \CL{} semantics.  E.g., the
calls in Figure~(\ref{fig:keyword-calls}) yield 42 as expected.  On
the contrary, the implementation does not treat \code{\&rest}
arguments in a special way (i.e., they are not \emph{thunked}), this
is because there is no way to access the list forming machinery in
\CL{} when \code{\&rest} arguments are present; in a lazy piece of
code, the list in the \code{\&rest} argument will contain the actual
thunks generated as if by \code{delay}.

\begin{figure}
\hrulefill
\begin{alltt}
   (lazy:call (lazy (lambda (x &key (y (loop) y-supplied-p))
                       (if y-supplied-p y (+ x 21))))
              21)

   (lazy:call (lazy (lambda (x &key ((:y yy) (loop)))
                         (if x (+ x 21) yy)))
              21)

   (lazy:call (lazy (lambda (x &key ((:y yy) (loop)))
                       (if x (+ x 21) yy)))
              nil :y 42))
\end{alltt}
\hrulefill
\caption{\code{\&key} arguments are dealt with as expected.  The
  answer is always, as expected, 42.}
\label{fig:keyword-calls}
\end{figure}

\section{Conclusions}

\CLAZY{} is an a exercise in \CL{} style, which is also useful.  The
\CLAZY{} library shows how, at the price of introducing a special call
operator (\code{lazy:call}), it is possible to introduce \emph{normal
  order} or \emph{lazy} evaluation in \CL{}.  The extension has the
following desirable characteristics: (i) it does not require the
construction of a full blown interpreter implementing lazy evaluation,
and (ii) thanks to the \code{deflazy} macro it allows a programmer to
write code in the most natural way.  It is much more
difficult to achieve the same effect in any other language than \CL{},
even when the language has macros.  It is the under-the-hood
interaction of macros and \code{symbol-macrolet} that makes \CLAZY{}
possible.

Of course, once this basic machinery is in place, extra \CL{}
incantations can be made and reader macros put in place as desired.

\CLAZY{} is not perfect of course.  The main open issue to complete
the integration within the frame provided by \CL{} is to work out a
way to deal with \textsf{CLOS} methods.  One way to achieve this would
be to automatically define a method specializing on thunks for a given
generic function.  While this may work, it does open up typing
issues\footnote{\code{lazy:call} would need to know the actual
  resulting type of the argument expressions to meaningfully set up a
  discrimination for the underlying method.} that need to be worked
out in details before proceeding with a full blown proposal.

\nocite{CLHS}

\bibliographystyle{plain}
\bibliography{cl-references}

\end{document}